\documentclass[11pt,letterpaper]{article}

\usepackage[backend=biber,
style=alphabetic,
sorting=nyt,
minalphanames=3,
maxbibnames=99]{biblatex}
\usepackage{amsmath,amsthm,amsfonts,amssymb}
\usepackage{hyperref}
\usepackage{graphicx}
\usepackage[capitalize]{cleveref}

\usepackage{comment}

\usepackage[utf8]{inputenc} %
\usepackage[T1]{fontenc}    %

\usepackage{booktabs}       %
\usepackage{amsfonts}       %
\usepackage{nicefrac}       %
\usepackage{graphicx}       %
\usepackage{bm}

\usepackage{fancybox}
\usepackage{varwidth}
\usepackage{diagbox}
\usepackage{multirow}
\usepackage{subcaption}
\usepackage{amssymb}
\usepackage{tikz} 
\usepackage{amsmath}
\usepackage{mathtools}
\usepackage{amsthm}
\usepackage{xcolor}
\usepackage[margin=1in]{geometry}

\usepackage[linesnumbered]{algorithm2e}
\RestyleAlgo{ruled}
\newcommand{\nosemic}{\renewcommand{\@endalgocfline}{\relax}}%
\newcommand{\dosemic}{\renewcommand{\@endalgocfline}{\algocf@endline}}%
\let\oldnl\nl%
\newcommand{\nonl}{\renewcommand{\nl}{\let\nl\oldnl}}%

\newcommand{\bigO}{\mathcal{O}}
\newcommand{\calG}{\mathcal{G}}

\newcommand{\calH}{\mathcal{H}}

\newcommand{\polylog}{\textit{polylog}}
 \newtheorem{theorem}{Theorem}[section]

\begin{document}

\begin{flushleft}
	\huge\bf
	Maintaining a Bounded Degree Expander in Dynamic
	Peer-to-Peer Networks
\end{flushleft}
\smallskip
\begin{flushleft}
	\setlength{\parskip}{3pt}

	\textbf{Antonio Cruciani} · Aalto University

\end{flushleft}	

\smallskip

\paragraph{Abstract.}

We study the problem of maintaining robust and sparse overlay networks in fully distributed settings where nodes continuously join and leave the system. This scenario closely models real-world unstructured peer-to-peer networks, where maintaining a well-connected yet low-degree communication graph is crucial. We generalize a recent protocol by Becchetti et al.~[SODA 2020] that relies on a simple randomized connection strategy to build an expander topology with high probability to a dynamic networks with churn setting. In this work, the network dynamism is governed by an oblivious adversary that controls which nodes join and leave the system in each round. The adversary has full knowledge of the system and unbounded computational power, but cannot see the random choices made by the protocol. Our analysis builds on the framework of Augustine et al.~[FOCS 2015], and shows that our distributed algorithm maintains a constant-degree expander graph with high probability, despite a continuous adversarial churn with a rate of up to $\bigO(n/\polylog(n))$ per round, where $n$ is the stable network size. The protocol and proof techniques are not new, but together they resolve a specific open problem raised in prior work. The result is a simple, fully distributed, and churn-resilient protocol with provable guarantees that align with observed empirical behavior.

\thispagestyle{empty}
\setcounter{page}{0}
\newpage
\section{Introduction}

We study a simple \emph{dynamics} that builds and maintains sparse and well connected graph despite an \emph{adversarial churn} at each round.

As a motivating example, consider Peer-to-Peer (P2P) networks that are (for example) at the base of major blockchain systems such as Bitcoin and Ethereum in which the system provides decentralized communication and resilience against single points of failure. These overlays are governed by strict privacy rules restricting any leak of sensitive information about the peers. Compliance to such privacy rules essentially makes it difficult to employ any distributed protocol that explicitly uses standard message passing techniques. For example (say in the P2P Bitcoin Network), it is not possible to use random walks to ``uniformly spread'' the node IDs (in this case IP Addresses) on the network and provide peers with a fresh set of uniform IDs from which to sample new neighbors. Indeed, after an initial bootstrap phase in which the nodes rely on DNS seeds for node discovery, nodes running the Bitcoin-core implementation turn to a decentralized policy to rebuild their neighborhood when their degree drops below the configured threshold. Each node has a minimum and a maximum number of neighbors that they must have (respectively $8$ and $125$, in the default configuration) and it locally stores a large list of IP addresses of active nodes. Every time the number of current neighbors of a node drops below the configured minimum value, it tries to find new connections with nodes sampled from its list. Such a list, is periodically shared to its neighbors and updated with the lists received by the neighbors. In the long run each node samples its neighbors from a list that represent a sufficiently random subset of nodes of the network. This sampling-based mechanism, despite being decentralized and oblivious to the global structure of the network, enables the construction of a sparse and robust overlay that induce good expansion properties. In particular, Becchetti et al.~\cite{Becchetti_2020} formalized and analyzed a dynamic random graph model inspired by the Bitcoin protocol, showing that the resulting network converges rapidly to a constant-degree expander with high probability.

However, real-world P2P networks are inherently dynamic: nodes frequently join and leave the system over time. In this brief announcement, we address an open question posed by Becchetti et al.~\cite{Becchetti_2020} by adapting their RAES protocol to a dynamic setting with adversarial churn. Our solution builds directly on the analytical framework developed by Augustine et al.~\cite{Augustine_2015}, and shows that a RAES-style protocol can maintain expansion with high probability even under continuous adversarial churn. 
This work serves as a theoretical closure to earlier experimental studies of RAES variants under churn~\cite{Cruciani_2023}. While the protocol and techniques are not original--both are adapted from prior work--the analysis formally confirms the empirical behavior observed in simulations. The result is a modest but rigorous endpoint to a research line focused on the resilience of randomized overlay networks.
\section{Preliminaries}
A \emph{dynamic graph}\footnote{In this paper we will use dynamic-graph and dynamic-network interchangeably.} $\calG$ is a family of graphs $\calG =\{G_t=(V_t,E_t):t\in \mathbb{N}\}$. We call $G_t$ the \emph{snapshot} of the dynamic graph at time $t$. For a set of nodes $S\subseteq V_t$, the \emph{edge-expansion} of the snapshot $G_t$ is defined as the minimum over all set of nodes $S\subseteq V_t$ whose size is $0<|S|\leq n/2$  of the ratio between the number of edges leaving the set and its size. 

\subparagraph*{The Dynamic Network with Churn Model.}
We consider a synchronous dynamic network controlled by an \emph{oblivious} adversary. The adversary fixes a sequence of graphs $\calH = (H_1,H_2,\dots, H_t,\dots)$ with each $H_t$ defined on a vertex set $V_t$. The size of the vertex set is assumed to be stable. In other words, we have $|V_t|=n$ for all $t$. Furthermore, we assume that $|V_t\setminus V_{t+1}| = |V_{t+1}\setminus V_t| \in \bigO(n/\log^k n)$ where $k$ is a constant determined in the analysis. The node set $V_t$ is the set of nodes present in the network at round $t$. We assume that each node has a unique ID chosen from an ID space of size $\text{poly}(n)$. The degree of each graph $H_t$ is bounded by a constant $\delta$. For the first $B=polylog(n)$ rounds, the adversary is \emph{silent}, i.e., there is no churn, more precisely $H_1 = H_2 =\dots =H_B$. Such an initial period is called \emph{bootstrap phase}, and we can think of this phase as a period of stability during which the protocol builds and initial bounded degree expander graph. Subsequently, the network is said to be in the \emph{maintenance phase} during which it can experience churn in the sense that a large number  of nodes might join and leave dynamically at each step. During the maintenance phase, we require that any new node $v$ that enters the network at round $t$ must be connected in $H_t$ to at least one pre-existing node $v$, i.e., if $u\in V_t\setminus V_{t-1}$, then $u$ must be connected to some $v\in V_{t-1}\cap V_i$ in round $t$. The adversary must ensure that the degree bound $\delta$ is not violated. We only require that a new node is connected to a pre-existing node in its very first round upon entering the network. Secondly, after the bootstrap phase we do not require $H_t$ to be connected. Indeed, the only required edges in $H_t$ are the edges connecting new nodes to pre-existing ones. This is a standard requirement in the DNC model (see, for example,~\cite{Augustine_2012,Augustine_2015}) and prevents the adversary to create disconnected clusters. Our goal is to design a simple protocol that maintains a graph process
$\calG = \{G_t=(V_t,E_t)\}_{t\geq 1}$
over time such that each $G_t$ for $t\geq B$, has good expansion while maintaining a constant degree bound between $d$ and $\Delta$ with high probability. We assume that, at time $t$, all the nodes have the ability of sampling from a uniform distribution over the vertex set at time $t$.

\noindent\textbf{Remark.}  Although the assumption that a node can pick
its neighbors uniformly at random among all nodes of the
network is unrealistic in many scenarios, the edge creation processes in our model is reminiscent of
the way some unstructured peer-to-peer networks such as the Bitcoin Network maintain a ``random'' topology.

\subparagraph*{The RAES Protocol.}
\textit{RAES (Request a link, then Accept if Enough
Space)}~\cite{Becchetti_2020} is a random graph model defined by
three parameters $n \in \mathbb{N}, d \in \{1, \dots, n-1\}, c > 1$, in which
each one of $n$ nodes has degree exactly $d$ and degree at most $cd=\Delta$.
The random graph is generated according to the discrete random
process described in Algorithm~\ref{algo:RAES}. The process terminates when all nodes have
degree in $[d,\Delta]$.

\begin{algorithm}[htb!]
	\caption{Overview of the RAES protocol.}
	\label{algo:RAES} 
	
    Let $G = (V,\{\emptyset\})$.
    
    \ForEach{$t\geq 0$}{
        \textbf{Phase 1 (reconnection):} Each node $u$ with degree $d(u)<d$, picks $d-d(u)$ new neighbors uniformly at random.
		
		\textbf{Phase 2 (degree adjustment):} Each node $u$ with degree $d(u)>\Delta$ selects $d(u)-\Delta$ neighbors $u$ uniformly at random and drops the connection with them.
    }
	
\end{algorithm}

The RAES model can be seen as a simplified version of the network-formation
process implemented in Bitcoin-core~\cite{Nakamoto_2009,Bitcoin_core}.  
The protocol converges in $\bigO(\log n)$ rounds to an expander graph with high probability~\cite{Becchetti_2020}. However, this is not guaranteed to happen when nodes continuously join and leave the network.

\section{Our Contribution}
We address the problem of defining a variation of the RAES protocol that can build and maintain a dynamic expander graph despite an adversarial churn of $\bigO(n/\log^k n)$ at each round.  Our construction relies on periodic randomized neighbor refreshing and pruning of high-degree nodes, and remains lightweight in terms of memory and messaging overhead. Our solution tackles one of the open problems in~\cite{Becchetti_2020} by building directly on the framework developed by Augustine et al.~\cite{Augustine_2015}.
\subparagraph*{The D-RAES Protocol.} We now introduce D-RAES, a dynamic extension of the RAES protocol that is resilient to a continuous adversarial churn of up to $\bigO(n/\polylog(n))$ nodes per round. The execution of the protocol is divided into two main phases: a bootstrap phase and a maintenance phase. During the bootstrap phase, the goal is to construct a bounded-degree expander graph. In the maintenance phase, the protocol continually repairs and adapts the network in response to churn events. The bootstrap phase begins with a graph o $n$ nodes and no edges. We run the RAES protocol (Algorithm~\ref{algo:RAES}) for $B = \bigO(\log n)$ rounds. As shown in~\cite{Becchetti_2020}, this results in an expander graph whp. (see Theorem 2.1 in~\cite{Becchetti_2020}). After this phase, the adversary is allowed to churn nodes, and each node executes a maintenance cycle composed of three phases (Algorithm~\ref{algo:overview}).

\begin{algorithm}[htb!]
	\caption{Overview of the D-RAES protocol. }
	\label{algo:overview} 
	
    Let $G$ be the graph obtained after the Bootstrap Phase.

	\ForEach{$t> B$}{

    \textbf{Phase 1 (refresh neighbors):} With probability $1/\log^k n$ each node $u$ with degree $d(u)\in [d,\Delta]$ drops all its neighbors.
        
		\textbf{Phase 2 (reconnection):} Each node $u$ with degree $d(u)<d$, picks $d-d(u)$ new neighbors uniformly at random.
		
		\textbf{Phase 3 (degree adjustment):} Each node $u$ with degree $d(u)>\Delta$ selects $d(u)-\Delta$ neighbors $u$ uniformly at random and drops the connection with them.
	
	} 	
\end{algorithm}

The high level idea of the protocol is that each node in the network seeks to maintain a bounded degree graph by keeping a degree between the two parameters $d$ and $\Delta$ with $d<\Delta$. This simple procedure mimics the behavior of the Bitcoin Network formation process~\cite{Bitcoin_core}. We require that nodes in the network that were not churned by the adversary they must refresh their neighbor list with probability $1/\polylog(n)$. This is fundamental to avoid the oblivious adversary incrementally building bad structures that can gradually degrade and ultimately destroy the expansion of the graph. Thus, we require that on average, nodes must refresh their neighborhood every $\polylog(n)$ rounds. Indeed, the authors in~\cite{Cruciani_2023} experimentally showed that refreshing long-lasting edges in the network might help the process to recover after a churn. The protocol uses a simple mechanism to understand when it is time to reconnect: when the number of neighbors falls below $d$ it infers that is not well-connected anymore and it tries to find new random neighbors from the network. In addition, the protocol preserves bounded degrees by ``pruning'' neighbors of nodes that exceed $\Delta$.

As mentioned before, the main intuition behind the protocol is that it tries to maintain a graph that resembles a random-bounded degree graph in every round whp. Thus, in addition to reconnection that are caused due to lost edges, in every round every node with probability $\Theta(1/\polylog(n))$, simply deletes all its edges and reconnects again.%
This refreshing step continuously creates new ``random enough'' edges in the network, which is useful in showing expansion properties in every round. Moreover, it is possible to show that if we do not include such a phase in the protocol (or an analogous step) it is impossible to maintain a constant degree expander graph under adversarial churn.

\begin{theorem}
    Assume to have an expander graph $G_t$ after the bootstrap phase and that the maintenance protocol does not include any edge refreshing phase. Then, there is an adversarial strategy that (1) maintains a constant degree graph in which each node has degree in $[d,\Delta]$ and (2) the graph is not anymore an expander graph.
\end{theorem}
We show an high-level overview of the proof. The core idea is that there is an adversarial strategy that gradually erodes the well connected part of the graph while maintaining a bounded-degree graph. Without the edge-refresh phase (and thus using the RAES protocol as it is in Algorithm~\ref{algo:RAES}) the adversary can: churn out some nodes from the graph and immediately insert $n/\polylog(n)$ nodes by connecting these new nodes to a low-expansion region and avoid touching the core that keeps being and expander. However, over multiple rounds the well-connected part of the graph shrinks due to the churn while the weakly connected fringe grows. Since we do not have any mechanism to re-randomize the bad edges we have that the expansion collapses even though the graph is still a bounded-degree graph.

The graph produced after each iteration of Algorithm~\ref{algo:overview} (lines 2-6), is the result of the interaction between two entities: the oblivious adversary and Algorithm~\ref{algo:overview}. To analyze the distributed protocol behavior under churn, we use the same approach of Augustine et al.~\cite{Augustine_2015} and we define a family of processes whose behavior is: (1) the adversary's move; (2) the actions by the nodes; and, (3) the random bits available to the process itself. As in~\cite{Augustine_2015}, we refer to a member of this family of processes as the $\Upsilon$-process.
\subparagraph*{Overview of the process.} 
We present a high-level description of the process and intuitions behind its analysis and  in the full version of this work we will provide a complete description and analysis of the process. The main idea is that the process captures both, the distributed protocol's and the adversary's actions by first applying the adversarial churn to the graph and then steps (1-3) in Algorithm~\ref{algo:overview} at each round. We have that the $\Upsilon$-process starts from a bounded degree expander graph that was created during the bootstrap phase. At each iteration it generates a new graph $G_t =(V_t,E_t)$, given the graph at $t-1$. The process ensures that nodes with a degree less than $d$ and a life-span greater than $\Omega(\log n)$ will gain at least once a degree between $d$ and $\Delta$ in $\bigO(\log n)$ rounds. Moreover, to argue that at each round we have a large expander subgraph we have to take into account that: (1) the (adversarial) churn is $\bigO(n/\log^k n)$, (2) the number of nodes that choose to refresh their neighborhood is $\bigO(n/\log^k n)$ whp., and (3) the number of ``unlucky'' nodes that after the reconnection phase had degree exactly $d$ and that after the degree adjustment phase lost some neighbors (due to some highly connected node with degree greater than $\Delta$ dropping a random subset of their neighbors) is $\bigO(n/\log^{k} n)$ whp. By iteratively using the results by Bagchi et al. in~\cite{Bagchi_2006}, the fact that nodes sample from a uniform distribution over $V_t$ and that no node can have degree less than $d$ for $\Omega(\log n)$ consecutive rounds whp. we can argue that at the end of each iteration $t$, the $\Upsilon$ process produces a graph $G_t$ that contains a large core with good expansion properties. This result implies that the D-RAES protocol maintains a graph with a $(n-o(n))$-sized constant expander subgraph in which all the nodes have degree between $d$ and $\Delta$. Our main result, that will be proved in the full version of this work, can be summarized as follows. 
\newpage

\begin{theorem}[Main Theorem]
	Despite an adversarial churn rate $\bigO(n/\log^k n)$ for $k\geq 1$, 
	the D-RAES protocol maintains a dynamic graph $\calG = \{G_t=(V_t,E_t)\}_{t\geq 1}$ in which each snapshot contains a large $n-o(n)$-sized expander graph with edge expansion at least $\alpha$ for a number of rounds that is at least $n^c$ for and arbitrarily big $c\geq 1$, with high probability.
\end{theorem}
\section{Conclusion}

We presented a fully distributed and lightweight protocol for maintaining a constant-degree expander graph despite a continuous adversarial churn of up to $\bigO(n / \polylog(n))$ nodes per round. Despite this strong adversary, our protocol ensures that the resulting overlay graph remains an expander with high probability over time.
Moreover, this work demonstrates that robustness and good connectivity can be achieved under far weaker assumptions, without relying on global knowledge or heavy coordination among nodes. 
Our analysis suggests that, if nodes in the network periodically refresh their neighborhood, the Bitcoin Network creation Protocol can maintain a good expander graph that preserves a good edge expansion against heavy adversarial churn by simply executing a local dynamics in which nodes do not need any information about the overall network structure. Finally, our result closes an open problem in~\cite{Becchetti_2020} where the authors suggestested to study their protocol under churn settings.

\printbibliography

@inproceedings{Augustine_2015,
  author       = {John Augustine and
                  Gopal Pandurangan and
                  Peter Robinson and
                  Scott T. Roche and
                  Eli Upfal},
  title        = {Enabling Robust and Efficient Distributed Computation in Dynamic Peer-to-Peer
                  Networks},
  booktitle    = {{IEEE} 56th Annual Symposium on Foundations of Computer Science, {FOCS}
                  2015, Berkeley, CA, USA, 17-20 October, 2015},
  publisher    = {{IEEE} Computer Society},
  year         = {2015}
}

@inproceedings{Becchetti_2020,
  author       = {Luca Becchetti and
                  Andrea Clementi and
                  Emanuele Natale and
                  Francesco Pasquale and
                  Luca Trevisan},
  title        = {Finding a Bounded-Degree Expander Inside a Dense One},
  booktitle    = {Proceedings of the 2020 {ACM-SIAM} Symposium on Discrete Algorithms,
                  {SODA} 2020, Salt Lake City, UT, USA, January 5-8, 2020},
  publisher    = {{SIAM}},
  year         = {2020}
}

@inproceedings{Cruciani_2023,
  author       = {Antonio Cruciani and
                  Francesco Pasquale},
  title        = {Dynamic graph models inspired by the Bitcoin network-formation process},
  booktitle    = {24th International Conference on Distributed Computing and Networking,
                  {ICDCN} 2023, Kharagpur, India, January 4-7, 2023},
  publisher    = {{ACM}},
  year         = {2023}
}

@article{Bagchi_2006,
  author       = {Amitabha Bagchi and
                  Ankur Bhargava and
                  Amitabh Chaudhary and
                  David Eppstein and
                  Christian Scheideler},
  title        = {The Effect of Faults on Network Expansion},
  journal      = {Theory Comput. Syst.},
  year         = {2006}
}

@inproceedings{Augustine_2012,
  author       = {John Augustine and
                  Gopal Pandurangan and
                  Peter Robinson and
                  Eli Upfal},
  title        = {Towards robust and efficient computation in dynamic peer-to-peer networks},
  booktitle    = {Proceedings of the Twenty-Third Annual {ACM-SIAM} Symposium on Discrete
                  Algorithms, {SODA} 2012, Kyoto, Japan, January 17-19, 2012},
  publisher    = {{SIAM}},
  year         = {2012}
}

@article{Nakamoto_2009,
  author = {Nakamoto, Satoshi},
  title = {Bitcoin: A Peer-to-Peer Electronic Cash System},
  url = {http://www.bitcoin.org/bitcoin.pdf},
  year = 2009
}

@misc{Bitcoin_core,
author={Bitcoin Core},
title = {Bitcoin Core P2P Network.
},
url = {https://en.wikipedia.org/wiki/Bitcoin_Core}
}
\newpage

 \end{document}